\documentclass[%
 aip,
 amsmath,amssymb,
 reprint,%
]{revtex4-2}

\usepackage{graphicx}%
\usepackage{dcolumn}%
\usepackage{bm}%

\usepackage[utf8]{inputenc}
\usepackage[T1]{fontenc}
\usepackage{mathptmx}
\usepackage{etoolbox}

\makeatletter
\def\@email#1#2{%
 \endgroup
 \patchcmd{\titleblock@produce}
  {\frontmatter@RRAPformat}
  {\frontmatter@RRAPformat{\produce@RRAP{*#1\href{mailto:#2}{#2}}}\frontmatter@RRAPformat}
  {}{}
}%

\makeatother
\begin{document}

\preprint{AIP/123-QED}

\title[Inverse design of Mie resonators with minimal backscattering]{Inverse design of Mie resonators with minimal backscattering}

\author{Vladimir Igoshin}
\author{Alexey Kokhanovskiy}
\author{Mihail Petrov}

\affiliation{School of Physics and Engineering, ITMO University}

 \email{m.petrov@metalab.ifmo.ru}

\date{\today}%

\begin{abstract}
Manipulation and engineering of  light scattering by resonant nanostructures is one of the central problems in optics and photonics. In this work, we theoretically study the effect of suppressed back-scattering of dielectric nanoantenna. We employed covariance matrix adaptation evolution strategy to identify the geometries of circular dielectric structures with minimized backward scattering cross section. Zero back-scattering is achieved due to generalized Kerker effect and multipole cancellation condition. We found a set of geometries and shapes of the nanoantenna  having  back-scattering intensity close to zero. With help of clustering algorithms,  all the found geometries fall separated into  several groups according to their multipolar content. While the optical properties of scatterers in each group were similar due to similar multipolar content, their shapes can be significantly different which stresses the ambiguity of free-form optimization problem. We believe that the obtained  results and found possible classes on generalized Kerker nanoantenna can help in designing nanophotonic structures such as antireflective metasurfaces. 
\end{abstract}

\maketitle

Modern nanophotonics is advancing towards more complex systems with an increasing number of controlling parameters\cite{kivshar2022rise, liu2023inverse}. For instance, changing the shape of metaatoms, which are the essential building blocks of optical metastructure, from  highly symmetric primitives such as cylinders and spheres to more sophisticated shapes immediately expands their optical functionality \cite{poleva_multipolar_2023,Alaee2015a,Terekhov2017,Kuznetsov2022Dec,CanosValero2023Aug, Wang2021Dec}. However,   increasing the complexity of metaatoms immediately results in a lack of analytical solutions and inevitable increase of computational complexity \cite{ohnoutek2022third}. Addressing this problem, development of  methods for inverse design of   metaatoms and metastructures with given optical properties and functionality becomes one of the topical fields of research \cite{Molesky2018,Bayati2020,estrada2022inverse, So2020May}.

Deep learning algorithms, particularly artificial neural networks, are recognized as universal approximators and are widely used for design tasks, having already shown remarkable results \cite{wiecha2019deep, Krasikov2022Mar}. These algorithms were applied to optimize optical properties of  metaatoms such as core-shell structures \cite{so2019simultaneous, estrada2022inverse}, nanoparticle arrays~\cite{Krasikov2022Mar} , thermal emitters~\cite{Kudyshev2020Jun}, including optimization the geometry of  free form objects \cite{li_machine_2024, Augenstein2023May, An2020Oct}.  Despite their success, there remains significant controversies and unanswered questions regarding the feasibility of using machine learning algorithms for design tasks ~\cite{wiecha2024deep}. One of the major challenges is obtaining a statistically significant dataset that adequately represents the problem. Alternative approach to inverse design of nanophotonic systems is based on evolutionary optimization methods \cite{volkov_non-radiative_2024, Wiecha2017Feb, Wiecha2019Sep}. In contrast to the machine learning algorithms, they do not require  preliminary generation of a dataset. Moreover, the advanced algorithms based on covariance matrix adaptation evolution strategy  (CMA-ES)   \cite{hamano2022cma} do not require  calculation of approximate gradients of the objective function either and can be used for  non-smooth, non-continuous, and nonlinear fitness functions   \cite{nomura2023cma}. 

In this work, we tackle the problem of finding optimal geometry of cylindrically symmetric dielectric scatterers with minimal backscattering cross-section. The effect of suppressed backscattering in nanophotonics is associated with Kerker effect \cite{Kerker1983, Liu2018May} and finds its applications in manipulating of nanoantenna directivity~\cite{Alaee2015Jun, Dobrykh2020Nov} and designing anti-reflecting coating \cite{Baryshnikova2016,Babicheva2017}. Suppression of backscattering has been already demonstrated in dielectric scatterers with basic  primitives shapes such as dielectric conical~\cite{Kuznetsov2022Dec, Terekhov2017} and spheroidal particles~\cite{Bukharin2022May},
dielectric cubes and pyramids~\cite{Terekhov2017}.
Limited number of parameters of these geometries primitives can be manually optimized, while address this problem from the free-form perspective.

\begin{figure}[t]
\centering
\fbox{\includegraphics[width=\linewidth]{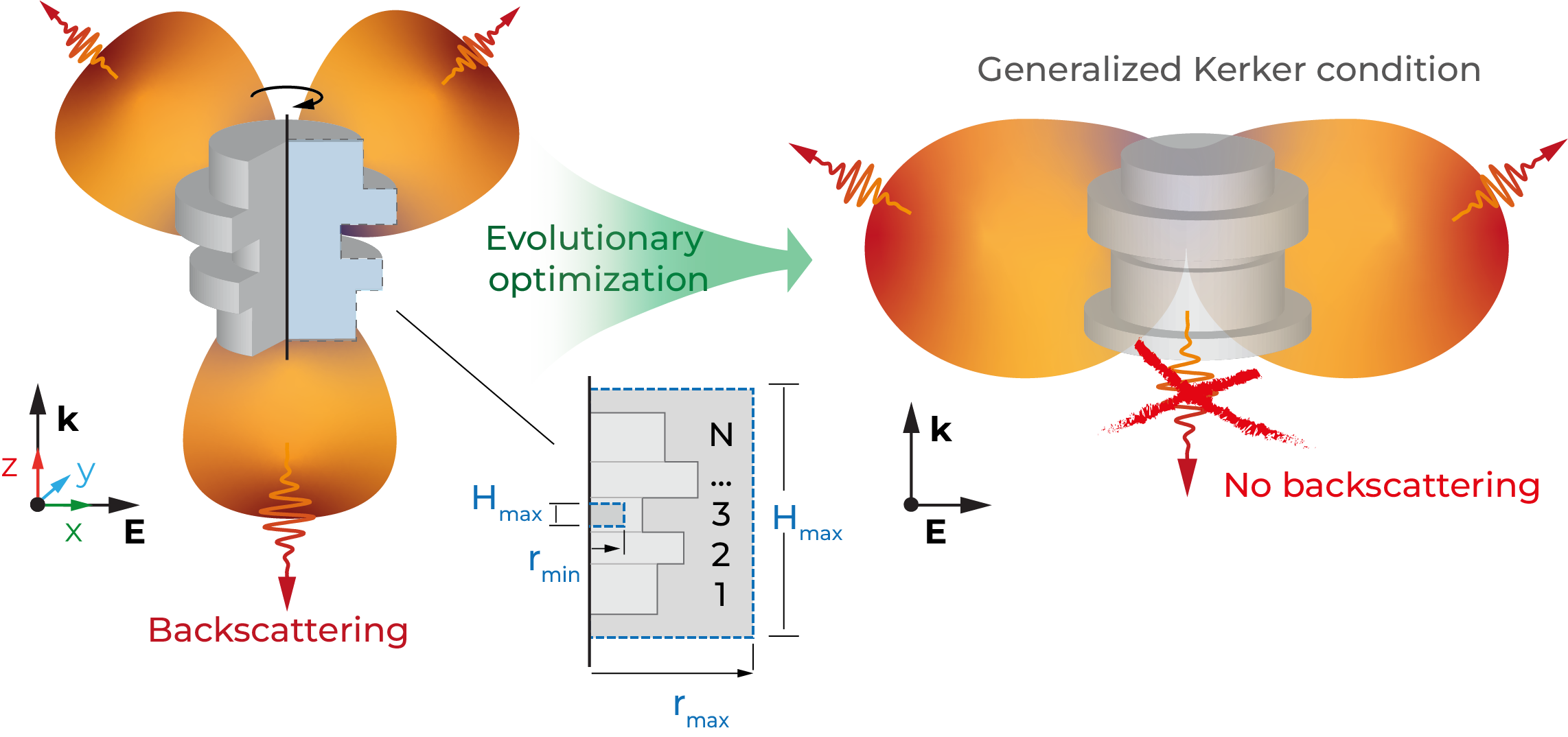}}
\caption{The main idea of this work is shown: evolutionary optimization algorithm finds the geometry of dielectric cylindrical nanoantennas consisted  of $N$ section having minimal scattering intensity on in the backward direction. The inset shows the geometry of the considered system before the revolution operation. The red dashed lines show the constraints: minimal and maximal size of the structure.}
\label{fig:1-geometry}
\end{figure}

\begin{figure}[t]
    \centering
    \fbox{\includegraphics[width=\linewidth]{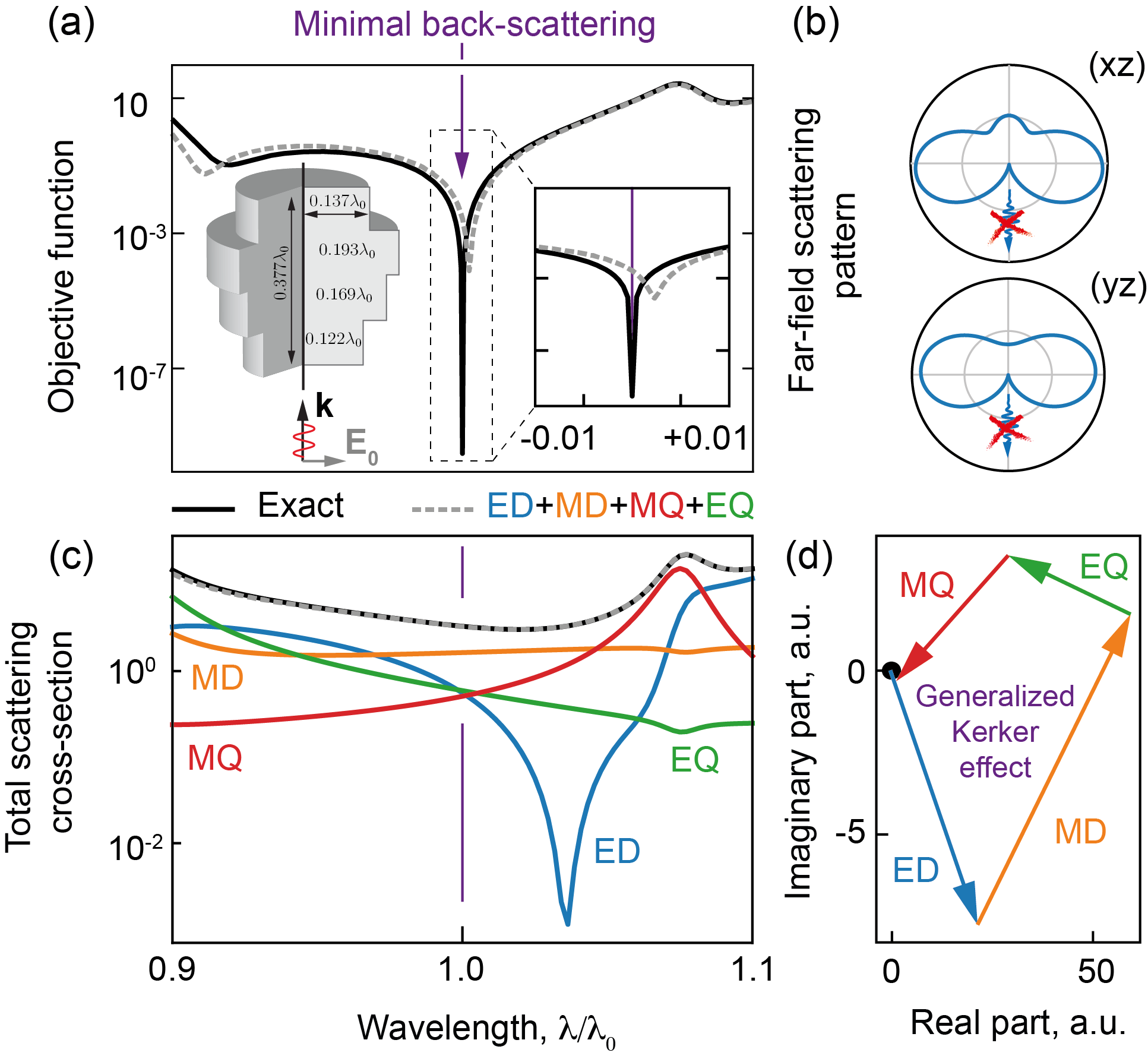}}
    \caption{Example of a scatterer with suppressed backward scattering.
    (a) Geometry of the scatterer and the spectrum of $\text{BCSn}$, showing a narrow dip at the wavelength $\lambda_0$.
    The inset compares the exact $\text{BCSn}$ value with the value calculated using multipole contributions.
    (b) Radiation patterns in the $xz$ and $yz$ planes.
    The patterns show forward and side radiation, with no backward radiation.
    (c) Total scattering cross-section and its decomposition into electric (magnetic) dipole $\text{ED}$ ($\text{MD}$), electric (magnetic) quadrupole $\text{EQ}$ ($\text{MQ}$) contributions.
    (d) Multipolar components of the backward radiating field in the complex plane.
    Each component is depicted as a vector, and their near-zero sum shows that the generalized Kerker condition is met.}
    \label{fig:2-example}
\end{figure}

Here, we utilize CMA-ES evolutionary optimization algorithm to find the geometry of high refractive index dielectric scatterer composed of $N$  sections, coaxially stacked cylinders, as shown in Fig.~\ref{fig:1-geometry}. By increasing the number of sections and, thus, the number of free parameters, we approach to the free-form geometry of the scatterer.    
We assume that the scatterers are placed in vacuum  surrounding. The refractive index of each cylinder is fixed to $n=4$ which is close to many dielectric materials such as  semiconductors in the visible and infrared ranges~\cite{Aspnes1983Jan} or for water~\cite{Kaatze1989Oct} and ceramics~\cite{Shamkhi2019May} for radiofrequncies.
The incident linearly polarized plane wave of wavelength $\lambda_0$ propagates along the axis ($z$-axis in Fig.~\ref{fig:1-geometry}) of the structure. We have a number of geometrical restrictions: while the radius $r_i$ of each cylinder is independent, the hight of each cylinder is set to $h_i=H/N$, where $H$ is the height of the whole structure. Thus,   for $N$ cylinders the structure is defined by $N+1$  parameters. The constraints for overall geometrical dimensions of the structure are also applied:
\begin{eqnarray}
    r_\text{min} \leq r_i \leq r_\text{max},\qquad
    H_\text{min} \leq H \leq H_\text{max},
\end{eqnarray}
where $r_\text{min} = 1/3\cdot\lambda_0 /n$, $r_\text{max}=1.2\cdot\lambda_0 /n$,
$H_\text{min}=0.4\cdot\lambda_0 /n$, $H_\text{max}=2\cdot\lambda_0 /n$.
These constraints naturally limit the size parameter and the number of possible Mie resonance that can be excited in the structure by a plane wave~\cite{Kuznetsov2016Nov} that simplifies the consideration and allows to identify the underlying physics beyond the observed backscattering suppression.

The  backward scattered radiation can be  characterized by the backward  scattering cross-section (BCS) quantity 
\begin{eqnarray}
    \sigma_\text{bw} = \lim_{r\to\infty} 4\pi r^2 \frac{|E_\text{bw}(r)|^2}{|E_0|^2},
\end{eqnarray}
where $E_\text{bw}$  is the amplitude of the scattered field in the far-field domain in the  direction opposite to the incident wave, $E_0$  is the amplitude of the incident plane wave~\cite{Alaee2015Jun}. As the objective function, we minimize the BCS at wavelength of $\lambda_0$ normalized over the minimal geometric cross section
\begin{eqnarray}
    \text{BCSn} = \frac{\sigma_\text{bw}}{\sigma_0^\text{min}}\to\min,
\end{eqnarray}
where $\sigma_0^\text{min} = \pi r^2_\text{min}$. We use CMA-ES optimization algorithm, which does not require extensive hyper-parameter optimization and suits for black-box optimization, when no preliminary information about the fitness function behavior is provided, as in our  case.
More details on the algorithm is provided in Supplementary Materials Sec.~1, while the details on simulation method are presented in Supplementary Materials Sec.~2. 

One of the found examples of scatterers with minimal BCSn  is shown in  the inset of Fig.~\ref{fig:2-example} (a).
The far-field scattering pattern of this  structure in $(xz)$- and $(yz)$-planes are depicted in Fig.~\ref{fig:2-example}~(b) showing dominant forward and sideways scattering while the BCSn is close to zero. The spectrum of BCSn  is presented in Fig.~\ref{fig:2-example}~(a) with a sharp minimum at the wavelength of $\lambda_0$ reaching the theoretical value of $10^{-9}$.

\begin{figure}
    \centering
    \fbox{\includegraphics[width=0.9\linewidth]{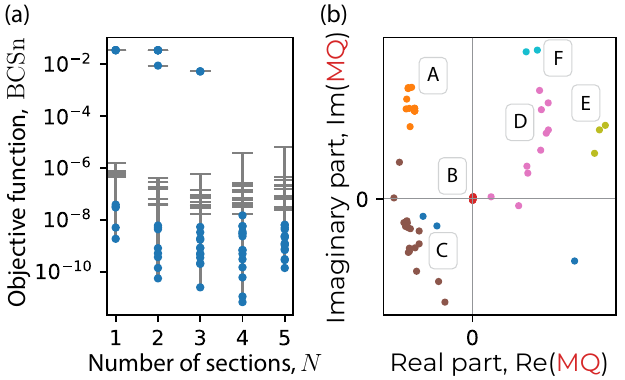}}
    \caption{(a) $\text{BCSn}$ values for all found solutions with near-zero backscattering.
    (b) Contribution of magnetic quadrupole (MQ) to the backward scattering field for all found solutions, represented in the complex plane.
    Dots with the same color represent solution within  the same cluster, each cluster marked by a number.
    Blue-colored dots are noisy samples and not included in any cluster.}
    \label{fig:3-results}
\end{figure}

\begin{figure*}
    \centering
    \fbox{\includegraphics[width=0.8\linewidth]{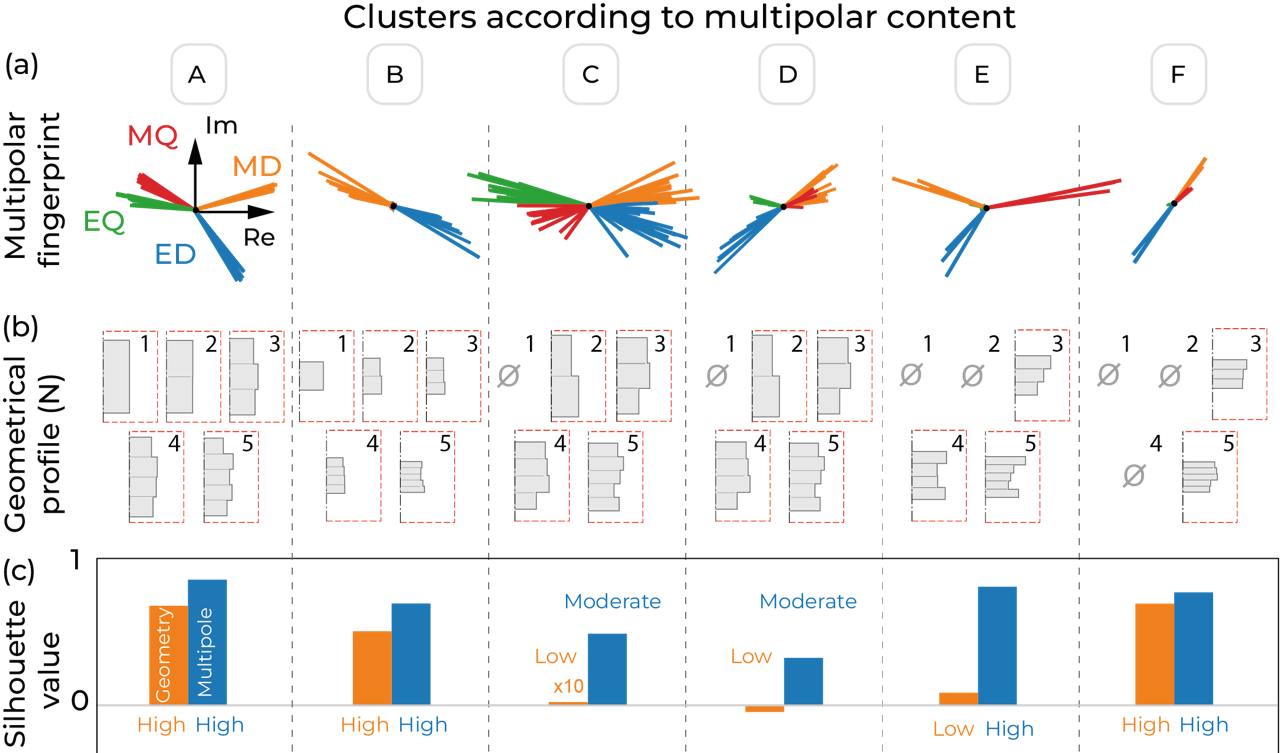}}
    \caption{Results of clustering analysis. Each row in the table corresponds to a single cluster, identified by a number. Column (a): multipolar contributions of all solutions within each cluster, represented in the complex plane. Column (b): examples of structures within each cluster for specific numbers of sections, ranging from $N=1$ to $N=5$. Column (c): Silhouette values (\text{Sv}) for each cluster, calculated using Euclidean distance in the space of multipolar contributions (Multipole) and the distance in the space of geometries (Geometry), as defined by~\eqref{eq:geom-distance}.
    Low, moderate, and high are related to cohesion of the cluster in the corresponding space.
    }
    \label{fig:4-clusters}
\end{figure*}

The  physical mechanism beyond the   observed BCSn suppression can be clarified with help of the  multipole theory which is a   powerful tool in electromagnetics \cite{Raab2005, poleva_multipolar_2023, Frizyuk2019Feb, Gladyshev2020Aug, Sadrieva2019Sep} and nanophotonics ~\cite{Liu2020, Kuznetsov2022Dec, Miroshnichenko2015Aug, Evlyukhin2011Dec}. Here, we will work in the basis of Cartesian multipoles for simplicity~\cite{Evlyukhin2016Nov, Mun2020May}.   Due to the size restrictions of our geometry, there are dominant contributions of the first  multipoles up to magnetic and electric quadrupoles only. Indeed, the scattering spectrum in  Fig.~\ref{fig:2-example} (c) shows partial cross sections correspondent to $x(y)$-components of electric (magnetic) dipoles, and $xy(yz)$-components of the electric (magnetic) quadrupoles  which being summed give the total scattering cross section with high accuracy. The other components of dipoles and quadrupoles are not excited due to the symmetry of  the problem. Thus,  the backward scattering field can be expanded in terms of the named multipoles~\cite{Kuznetsov2022Dec, Alaee2015Jun}
\begin{equation}
    E_{\text{bw}} = \frac{k^2}{4\pi\varepsilon_0}\frac{e^{ikr}}{r}\left(\underbrace{p_x}_{\text{ED}} + \underbrace{\frac{m_y}{c}}_{\text{MD}} \underbrace{- i\frac{k}{6}Q_{xy}}_{\text{EQ}} \underbrace{- i\frac{k}{6c}M_{yz}}_{\text{MQ}} + \dots\right),
    \label{eq:back-field}
\end{equation}
where the  multipolar moments are computed in accordance with Ref.~\cite{Alaee2018Jan}.
The real and imaginary parts of multipole field components in Eq.~\eqref{eq:back-field} at the wavelength $\lambda_0$ are shown in Fig.~\ref{fig:2-example}~(d) with arrows in a complex plane representing  electric dipole ($\text{ED}$), magnetic dipole ($\text{MD}$), electric quadrupole ($\text{EQ}$), and magnetic quadrupole ($\text{MQ}$) components.
One should note that the sum of these arrows  appears to be  close to zero that  corresponds to the to the so-called generalized Kerker condition~\cite{Kuznetsov2022Dec, Liu2018May, Alaee2015Jun}. Compared to the standard Kerker effect where only the sum of dipole components cancel each other, here all four components contribute to zero backscattering. At the same time,  it is not exact zero, although it  explains the origin of the appeared BCSn suppression. Fig.~\ref{fig:2-example}~(a) shows that the BCSn computed with the four multipoles only has minimum near the target wavelength $\lambda_0$, however neither its value nor the exact spectral position do not coincide with the numerically obtained ones. In order to get the exact matching, one should  consider contribution  of higher order multipoles, which may be unreasonable for the proposed physical interpretation.

Next, we addressed the  connection between the number of sections in the structure and the  obtained minimal value of BCSn. Increasing the number of sections, one increases the degrees of freedom in scatterer's geometry.  Fig.~\ref{fig:3-results}~(a) shows  $\text{BCSn}$ for all found minimal solutions depending on the number of sections $N$.
To our surprise, increasing the number of layers does not significantly decrease minimum value of $\text{BCSn}$ despite the growth of the optimization parameters number.
The typical minimal $\text{BCSn}$ values do not drop below the value of $10^{-9}$ for the considered values of $N$.
The near-zero $\text{BCSn}$ values are unstable and change dramatically even for small parameter perturbations.
This instability is illustrated with the error bars. More details on error calculation is provided are Supplementary Materials Sec.~3. The  reason to this instability and large error margins is related to numerical values  of drastically different scales, when  near-zero $\text{BCSn}$ is computed. 

Finally, we would like to turn to classification of the found scatterers according to their multipolar content. We visualized all found optimal solutions in  Fig.~\ref{fig:3-results}~(b) in a complex plane defining magnetic quadrupole (MQ) contribution in backward radiating field 
\begin{equation}
    \text{MQ} = - i\frac{k}{6c}M_{yz}.
\end{equation} %
Here we need to note that the imaginary part of quadrupole field components and related partial extinction cross-sections can have arbitrary sign as the optical theorem does not put any restrictions on the sign of partial cross-sections in the case of non-orthogonal channels\cite{Miroshnichenko2015_1, krasikov_multipolar_2021}.
We have applied DBSCAN clustering algorithm from \texttt{scikit-learn} Python library~\cite{Ester1996, scikit-learn} to analyze the obtained  datasets with all four dipole and quadrupole contributions according to Eq.~\eqref{eq:back-field}. The found different clusters are shown with colours in Fig.~\ref{fig:3-results}~(b).
Details on clustering parameters are presented in Supplementary Materials Sec.~4.

The obtained clusters are  visualized according to their multipolar content in the first row of Fig.~\ref{fig:4-clusters}. Colored arrows correspond to complex  field amplitude for different multipoles. Each cluster contains many solutions with relatively close multipolar content and, therefore, radiation pattern.
Radiation patterns  for each cluster are shown in Supplementary Materials Sec.~4. We need to stress that the found multipolar clusters form different types of generalized Kerker regimes which can be hardly predicted by means of any rigorous theory. Only the class B corresponds to standard dipole Kerker regime. However it is hard to conclude whether the obtained classes can be found for different geometries and morphology of scatterers.   

Each column in Fig.~\ref{fig:4-clusters}~(b) represents examples of optimized structures for specific number of sections from $N=1$  to $N=5$. One can notice that some clusters do not include structures with all values  of $N$. For instance, cylindrical structures $N=1$ enter only cluster A--B, and  structures with $N=2$ enter clusters A--C, while structures with $N=5$ enter all clusters. That supports an intuitive conclusion that the larger is the parameter space the larger number of clusters can be formed.

It is an intriguing  question how the geometrical form and multipolar content (far-field scattering pattern) are connected. To answer it, we have performed analysis of the ``distance" between different species in the parametric space  of their multipolar content and the distance in geometrical space using silhouette score $\text{Sv}$~\cite{Rousseeuw1987Nov} (see Supplementary Materials Sec.~4). For the multipolar components, we utilized Euclidean distance in 8--dimensional space of the real and imaginary parts of complex components.
For the geometries, we define a distance between two scatterers as follows
\begin{equation}
    d(i, j) = \int_{-\max(H_1, H_2)/2}^{\max(H_1, H_2)/2} |r_i(z) - r_j(z)| dz,
    \label{eq:geom-distance}
\end{equation}
where $r_i(z)$ is a shape curve defining the scatterer as a solid of revolution, centered at the geometrical center of the scatterer.
This induced distance compares the 2D shape of the geometric silhouettes of the structures shown in Fig.~\ref{fig:4-clusters}~(b). The obtained values are shown in Fig.~\ref{fig:4-clusters}~(c). The closer is Sv to 1, the smaller is the cluster distribution. One can see that while each cluster contains species with relatively close   multipolar content (Sv $\geq$ 0.5),  the geometrical Sv can be close to zero for some clusters (C, D, E). That basically means that each cluster contains examples of scatterers which have very close multipolar content   but significantly different geometrical shape. This is an another example proving that free-form geometry optimization of photonic structures is an ill-posed and ambiguous problem.

We have demonstrated the inverse design of metaatoms based on dielectric Mie resonators with cylindrical symmetry using the CMAES algorithm. The algorithm effectively identified resonator architectures made up of any number of sections that exhibit zero backscattering. The predicted back-scattering efficiency can be as low as $10^{-9}$ of geometrical cross-section. Tailoring the optical response of the resonator based on its architecture is, in principle, a multi-solution problem. Clustering the solutions according to the multipolar content reveals that there exist several classes satisfying generalized Kerker condition. Each class contains structures with the same   multipolar content and far-field radiation pattern but various geometry. Thus,  one can conclude that the multipolar expansion is a powerful tool for inverse design of nanophotonic structures with pre-designed optical properties.

\begin{acknowledgments}
 We thank  Pavel Ginzburg and Andrey Bogdanov for the discussions.  The work of A.K. was financially supported by the ITMO Fellowship Program. The work was supported by the Federal Academic Leadership Program Priority 2030.
 \end{acknowledgments}

\section*{Data Availability Statement} The data that support the findings of this study are available from the corresponding author upon reasonable request.

\bibliography{sample}

\end{document}


\maketitle

\section{Optimization  algorithm}

The main idea of the algorithm is following: each generation (step of the algorithm) a new set of candidate solutions is created.
The set is sampled in $N+1$~--~dimensional parameter space using the multidimensional normal distribution.
Then the fitness function is calculated for each candidate and a new mean vector and covariance matrix are calculated based on the obtained values.
The obtained mean vector and covariance matrix are used in the next generation to generate a new set of candidate structures.

We use Python-implemented CMAES library called \texttt{cmaes}~\cite{nomura2024cmaes}.

Initial step-size (standard deviation) is chosen $\sigma = 1/15 \lambda_0$.
The termination criteria was chosen such that standard deviation in the generation for each parameter is less than $1.67\cdot10^{-5} \lambda_0$ or the generations number exceeds 300.
Population size is different for each number of sections.
The number of total runs, population sizes, and mean number of iterations are presented in Table~\ref{tab:optimization}.

\begin{table}[htbp]
\centering
\caption{\bf Parameters of the optimization.}
\begin{tabular}{cccc}
\hline
$N$ & Total runs & Population size & Generations mean number \\
\hline
1 & 10 & 10 & 102.2\\
2 & 15 & 12 & 110.3\\
3 & 15 & 14 & 171.5\\
4 & 15 & 16 & 199\\
5 & 15& 18 & 236.5\\
\hline
\end{tabular}
  \label{tab:optimization}
\end{table}

\section{Numerical modeling}

We use the COMSOL Multiphysics{\textregistered} software and its wave optics module to calculate the scattering field of considered system~\cite{comsol}.
To drastically speed up calculations during the optimization process we use axial symmetry model with azimuthal numbers $m=-1,1$~\cite{Gladyshev2024Feb}.
We use spherical host domain with radius $\approx 5\lambda_0$ and perfectly matched layer with thickness $\approx 3\lambda_0$ to simulate infinite free space.
The element size of the mesh is chosen to be smaller than $\lambda_0/n/30$, where $n$ is refractive index of the meshed domain.
We calculate far-field using built-in Far-Field Domain.

\section{Error bars}

\begin{figure}[htbp]
\centering
\fbox{\includegraphics[width=\linewidth]{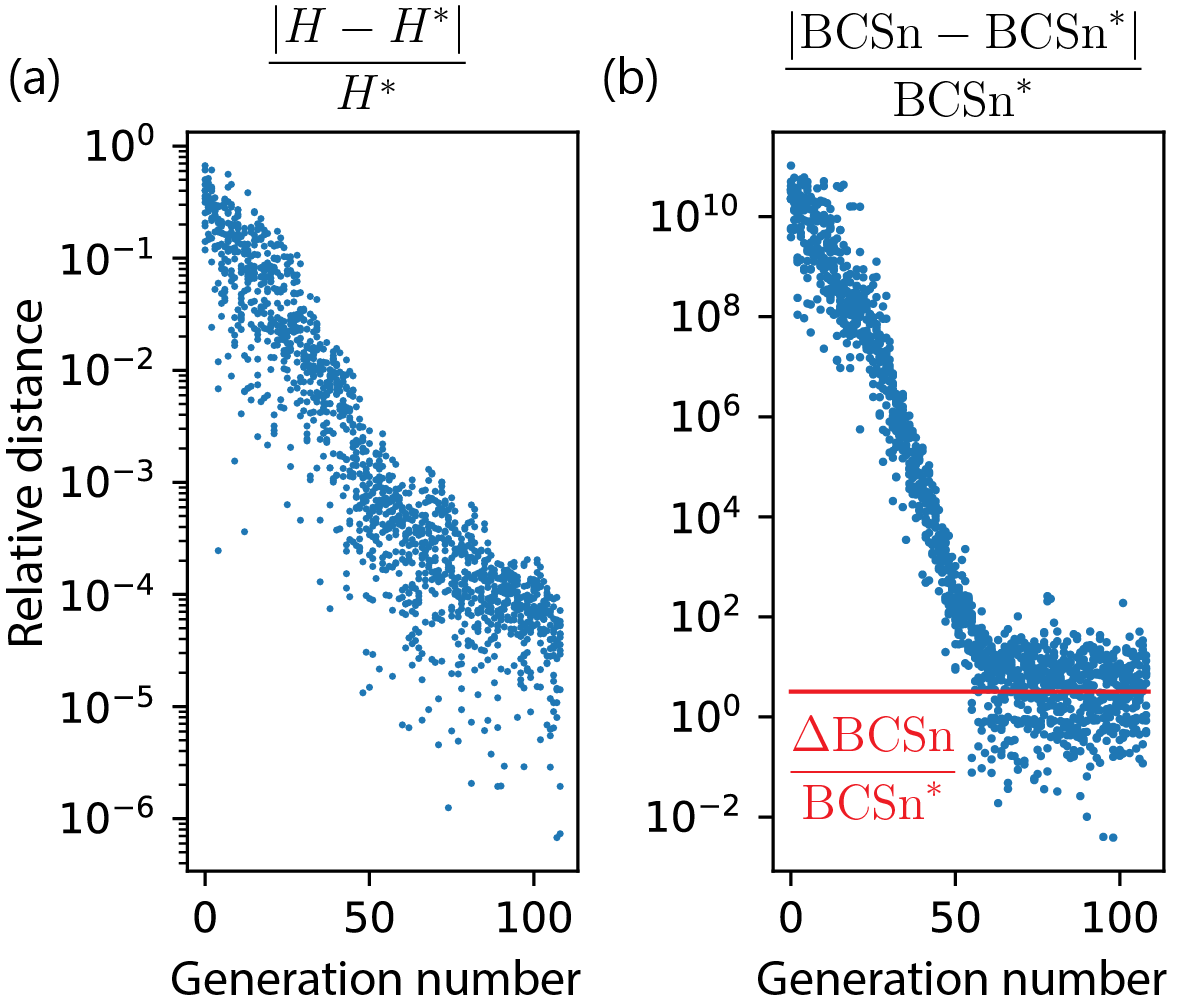}}
\caption{Relative distances between (a) the parameter $H$  and (b) $\text{BCSn}$ for each candidate at every iteration and the corresponding values $H^*$ and $\text{BCSn}^*$ for the best candidate.}
\label{fig:error_bars}
\end{figure}

In Fig.~\ref{fig:error_bars}~(a), the relative distance between the parameter $H$ for each candidate at every iteration and the corresponding value $H^*$ for the best candidate
\begin{equation}
    \frac{|H-H^*|}{H^*}    
\end{equation}
in the final generation is shown.
It is evident that the relative distance, and consequently the standard deviation (step-size), decreases with each iteration.
In contrast, the relative distance for $\text{BCSn}$, depicted in Fig.~\ref{fig:error_bars}~(b), stops decreasing after a certain generation and stabilizes around a fixed value.
These figures show that even small perturbations in the system parameter, and therefore small changes in the model mesh, can lead to significant variations in the $\text{BCSn}$ value.

The sensitivity of $\text{BCSn}$ does not correspond to a high derivative with respect to the parameter.
If it did, the relative distance of $\text{BCSn}$ would decrease as the parameter perturbations ($H$) decrease.
We interpret this sensitivity as indicative of a noisy fitness function landscape near very low 
$\text{BCSn}$ values due to computational errors.

To find the level the dependency stops decreasing, we calculate the average value of relative distance for $\text{BCSn}$ across all candidates over the last 10 generations.
This resulting value can be expressed as:
\begin{equation}
    \frac{\Delta\text{BCSn}}{\text{BCSn}^*},
\end{equation}
where $\Delta\text{BCSn}$ is absolute error of found $\text{BCSn}^*$ minimum.
The $\Delta\text{BCSn}$ is depicted as error bars in Fig.~3~(a) in the main text.
Minima associated with relatively high $\text{BCSn}$ values have smaller error bars, because these $\text{BCSn}$ values are large enough to avoid numerical issues, unlike the case with relatively small $\text{BCSn}$ values.

\section{Clusterization}

We choose minimum number of points equal 2 and $\varepsilon$ parameter for DBSCAN algorithm  such that silhouette score has a maximum value~\cite{Rousseeuw1987Nov}.

To calculate silhouette value $s_i$ for single data point $i$ of cluster $C_k$ one should introduce a distance between two data points $\operatorname*{d}(i, j)$.
The average distance between $i$ and all other data points in cluster $C_k$ is
\begin{equation}
    a_i = \operatorname*{mean}_{j\in C_k,\, j\neq i} \operatorname*{d}(i, j).
\end{equation}
The smallest distance between $i$ and any other cluster is
\begin{equation}
    b_i = \min_{C_t \neq C_k}\left( \operatorname*{mean}_{j\in C_t} \operatorname*{d}(i, j)\right).
\end{equation}
The silhouette value for single data point $i$ is defined by
\begin{equation}
    s_i = \frac{b_i - a_i}{\max(a_i, b_i)}.
\end{equation}
Now we introduce silhouette value for the cluster $C_k$, $\text{Sv}$
\begin{equation}
    \text{Sv} = \operatorname*{mean}_{i\in C_k} s_i.
\end{equation}
One can see that $-1\leq\text{Sv}\leq1$.
The larger value indicates better clustering and vice versa.

\begin{figure*}[htbp]
\centering
\fbox{\includegraphics[width=\linewidth]{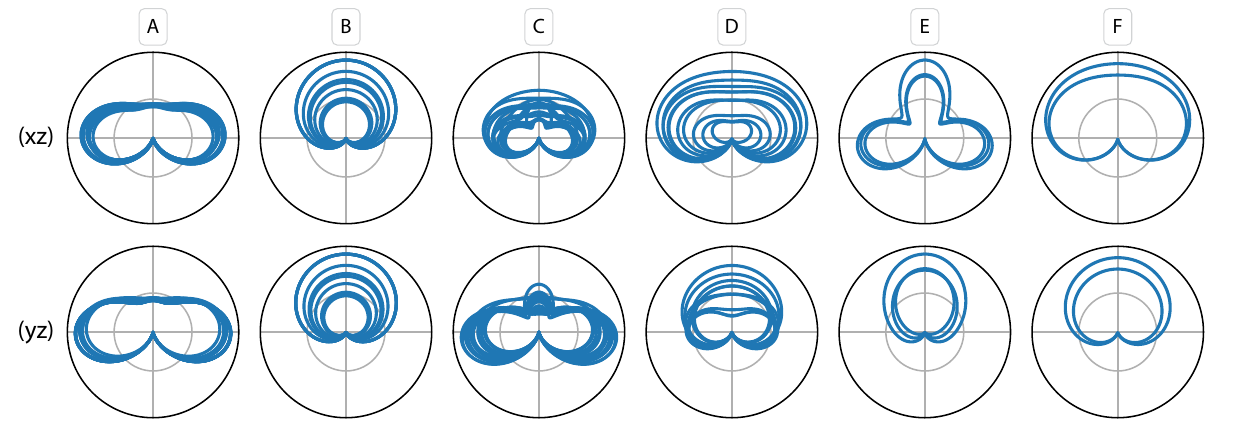}}
\caption{Radiation patterns for found clusters in xz and yz planes.}
\label{fig:6-patterns}
\end{figure*}

For clustering all minima found we use Euclidean distance between 8 features: real and imaginary parts of $\text{ED}$, $\text{MD}$, $\text{EQ}$, and $\text{MQ}$ .
All the features are normalized so that all of them lies in the range from -1 to 1.
The mean $\text{Sv}$ value over all the clusters of the resulting clustering is 0.59.
Radiation patterns for resulting clusters are shown in Fig.~\ref{fig:6-patterns}.

\bibliography{sample}